\documentclass[runningheads]{llncs}
\usepackage{graphicx}
\usepackage{hyperref}
\usepackage{amssymb}
\usepackage{lscape}
\usepackage[all,cmtip]{xy}
\usepackage{cancel}
\usepackage{bm}
\usepackage{multirow}
\usepackage{framed}
\usepackage{hyperref}
\usepackage{xcolor}
\usepackage{tikz, bbm}
\usepackage{tikz-cd}

\newcommand{\R}{\mathbb{R}}

\newcommand\Hthree[3]{\raisebox{-1pt}{\begin{tikzpicture}[box/.style={rectangle,draw}]
\draw (0,0) -- (1,0) -- (2,0);
\node[draw,box,inner sep=2.5pt,fill=white] at (0,0) {};
\node[draw,box,inner sep=2.5pt,fill=white] at (1,0) {};
\node[draw,box,inner sep=2.5pt,fill=white] at (2,0) {};
\node at (0.5,0.1) {{\tiny $5$}};
\node at (0,0) {$#1$};
\node at (1,0) {$#2$};
\node at (2,0) {$#3$};
\end{tikzpicture}}}
 
\newcommand\Hfour[4]{\raisebox{-1pt}{\begin{tikzpicture}[box/.style={rectangle,draw}]
\draw (0,0) -- (1,0) -- (2,0) -- (3,0);
\node[draw,box,inner sep=2.5pt,fill=white] at (0,0) {};
\node[draw,box,inner sep=2.5pt,fill=white] at (1,0) {};
\node[draw,box,inner sep=2.5pt,fill=white] at (2,0) {};
\node[draw,box,inner sep=2.5pt,fill=white] at (3,0) {};
\node at (0.5,0.1) {{\tiny $5$}};
\node at (0,0) {$#1$};
\node at (1,0) {$#2$};
\node at (2,0) {$#3$};
\node at (3,0) {$#4$};
\end{tikzpicture}}}

\newcommand\FourNodesDiagram[7]{\raisebox{-4pt}{\begin{tikzpicture}[box/.style={rectangle,draw}]
\draw (0,0) -- (1,0) -- (2,0) -- (3,0);
\node[draw,box,inner sep=2.5pt,fill=white] at (0,0) {};
\node[draw,box,inner sep=2.5pt,fill=white] at (1,0) {};
\node[draw,box,inner sep=2.5pt,fill=white] at (2,0) {};
\node[draw,box,inner sep=2.5pt,fill=white] at (3,0) {};
\node at (0.5,0.1) {{\tiny $#1$}};
\node at (1.5,0.1) {{\tiny $#2$}};
\node at (2.5,0.1) {{\tiny $#3$}};
\node at (0,0) {$#4$};
\node at (1,0) {$#5$};
\node at (2,0) {$#6$};
\node at (3,0) {$#7$};
\end{tikzpicture}}}

\begin{document}
\title{Studying Wythoff and Zometool Constructions using Maple
\thanks{BC supported by NSERC Discovery Grant. SW supported by NSERC USRA.}}
\author{Benoit Charbonneau\inst{1}\orcidID{0000-0001-7978-4466} \and\\
Spencer Whitehead\inst{2}\orcidID{0000-0001-8747-8186}}
\authorrunning{B. Charbonneau and S. Whitehead}
\institute{Department of Pure Mathematics, University of Waterloo, 200 University Avenue West,
Waterloo, Ontario, N2L 3G1, Canada.
\email{benoit@alum.mit.edu}\\
\url{http://www.math.uwaterloo.ca/~bcharbon} \and
Department of Pure Mathematics, University of Waterloo, 200 University Avenue West,
Waterloo, Ontario, N2L 3G1, Canada.
\email{snwhiteh@edu.uwaterloo.ca}\\
}
\maketitle              % typeset the header of the contribution
\begin{abstract}
We describe a Maple package that serves at least four purposes. First, one can use it to compute whether or not a given polyhedral structure is Zometool constructible. Second, one can use it to manipulate Zometool objects, for example to determine how to best build a given structure. Third, the package allows for an easy computation of the polytopes obtained by the kaleiodoscopic construction called the Wythoff construction. This feature provides a source of multiple examples. Fourth, the package allows the projection on Coxeter planes.

\end{abstract}

\section{Introduction} % (fold)
\label{sec:introduction}
As geometry is a very visual field of mathematics, many have found it useful to construct geometric objects, simple and complex, either physically or with the help of a computer. The Maple package we present today allows for both. First, the package allows for the automatic construction of many of the convex uniform polytopes in any dimension by the kaleidoscopic construction known as the Wythoff construction. Combined with the plotting capabilities of Maple exploited by our package, we therefore provide an extension of Jeff Weeks's KaleidoTile software \cite{KaleidoTile} to higher dimensions. This construction provides an impressive zoo of examples with which one can experiment on the second and most important aspect of our package, Zometool constructability.

\begin{figure}[h]\label{fig:Zome examples}
\centering
\includegraphics[height=3cm]{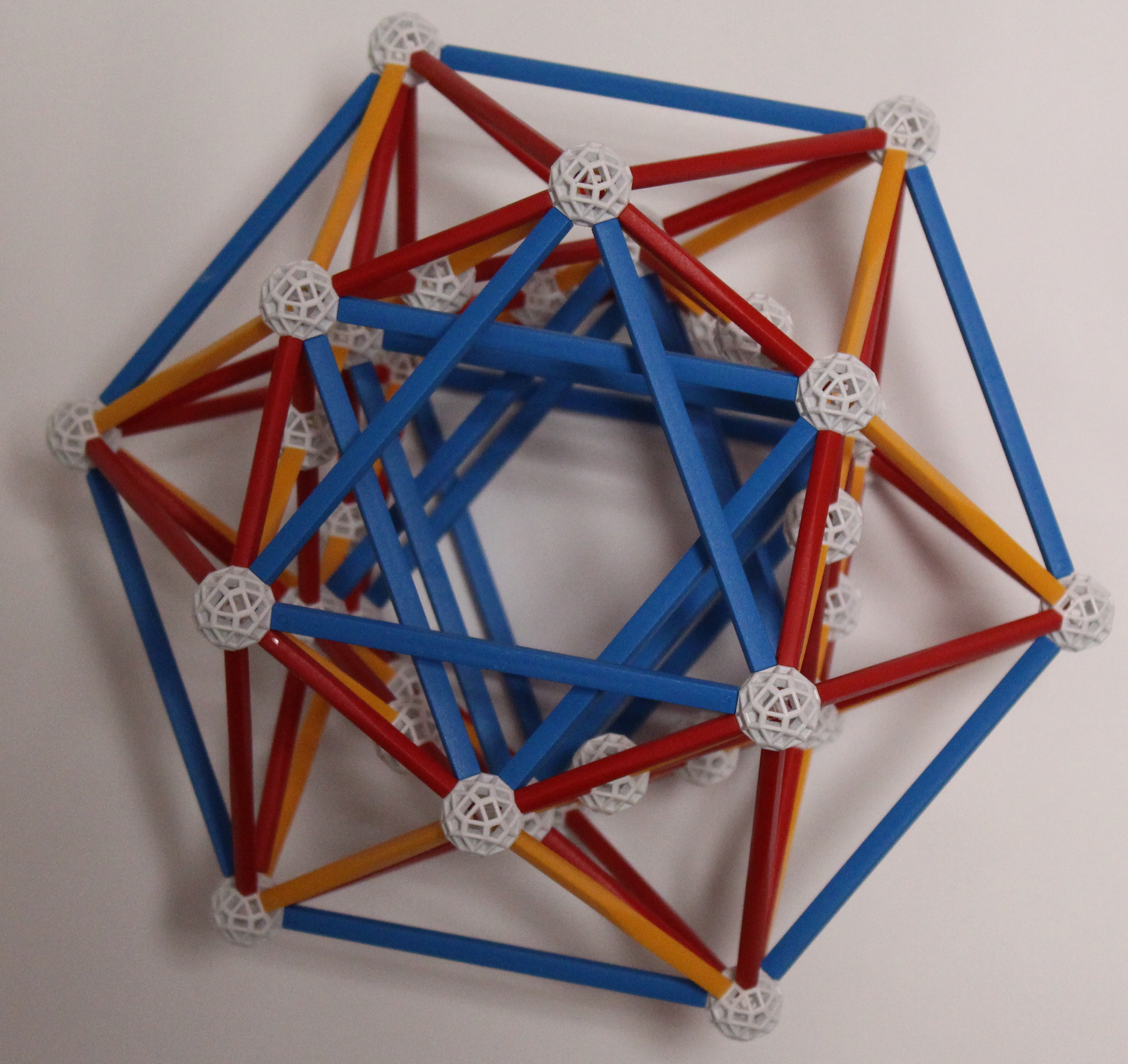}
\includegraphics[height=3cm]{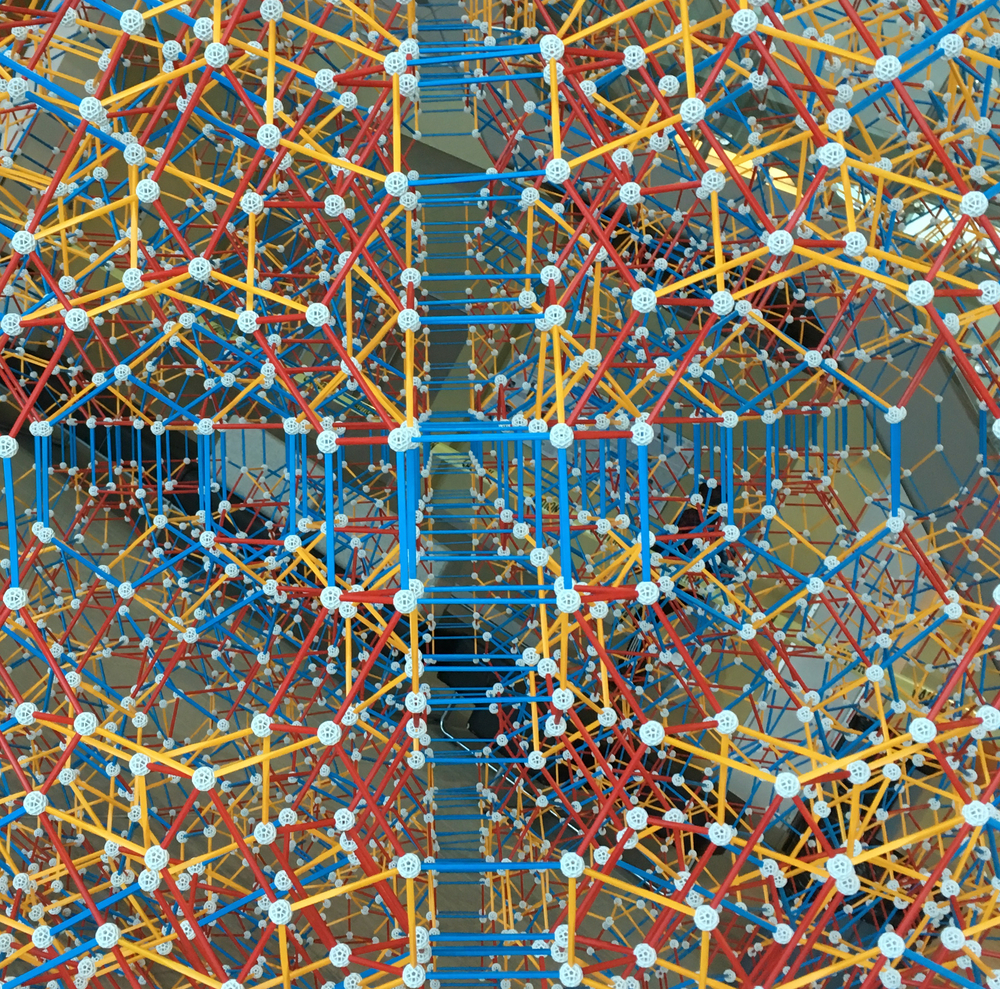}
\includegraphics[height=3cm]{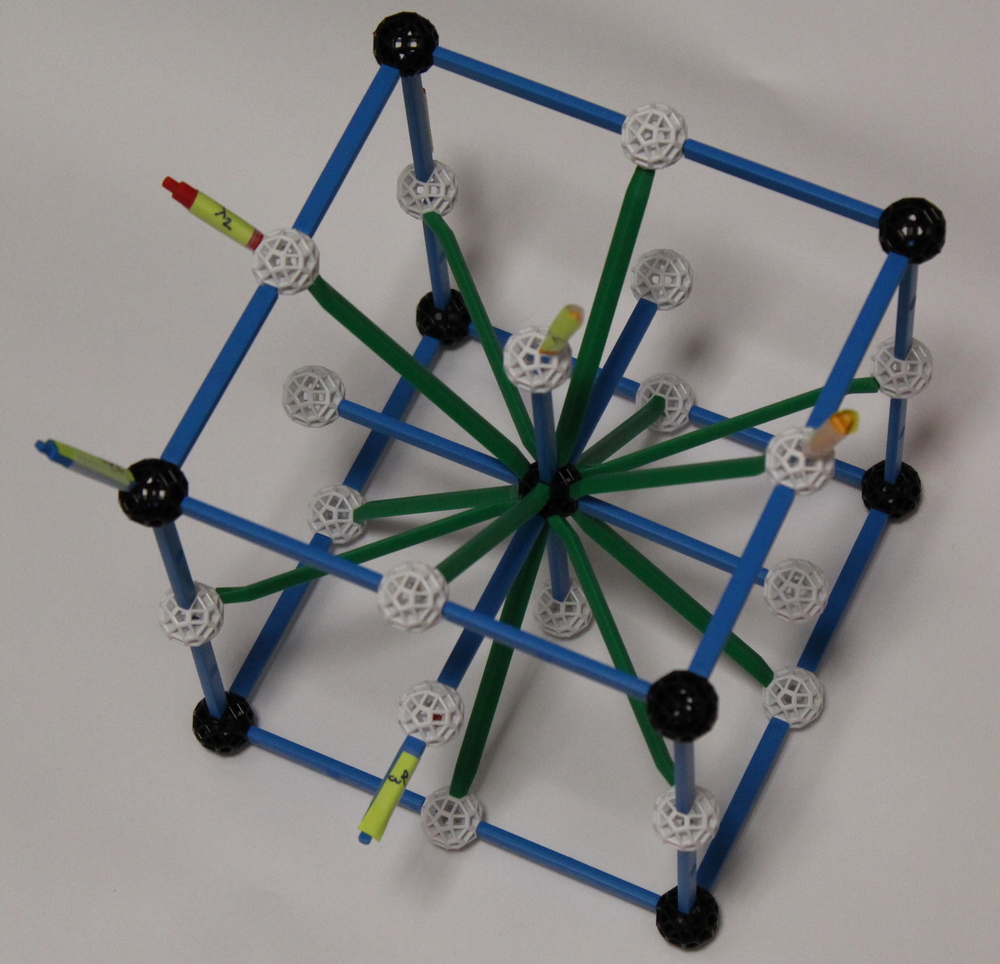}
\caption{A projection of the 24-cell, a close-up picture of the omnitruncated 120-cell along a 4-fold symmetry axis, and the root system of $B_3$.}
\end{figure}

The Zometool System as it currently exists was first released in 1992 by Marc Pelletier, Paul Hildebrant, and Bob Nickerson, based on the ideas of Steve Baer some twenty years earlier; see \cite{Zometool-about}. It is marketed by the company Zometool Inc.
As a geometric building block with icosahedral symmetry and lengths based on the golden ratio, it is able to construct incredibly rich structures with a very small set of pieces (see examples in Fig.~\ref{fig:Zome examples}). 
The Zometool System finds use today in many of the sciences:  modeling of DNA \cite{Voros-Zometool-B-DNA}, construction of Sierpi\'nski superfullerenes \cite{Chuang-Jin-Superfullerenes-Zome}, building models of quasicrystals (notably in the work of Nobel laureate Dan Shectman). 
Countless other mathematicians use the Zometool System for visualizing nuanced geometric structures easily (see for instance \cite{Hart-zome-polytopes-BarnRaising,Richter-zome-600-cell,Richter-Vorthman-Green-Quaternions-OctahedealZome}). A notable example is Hall's companion \cite{Hall-GeometryRootSystems} to the Lie algebra textbook \cite{Hall-textbook}. Many have used Zometool to teach or raise interest in mathematics \cite{Hart-Picciotto-ZomeGeometry,Hildebrandt-Zome-workshop}. 
Zometool is also an invaluable tool for educators, who find great success in using it to teach geometry in a hands-on fashion. 

Hart and Picciotto's book \emph{Zome Geometry} \cite{Hart-Picciotto-ZomeGeometry} allows one to learn about geometric objects, polygons, polyhedra, and polytopes and their projections in a hands-on fashion.
It contains instructions allowing one to build many projections. However, it can be difficult to verify that one is building the claimed structure from such instructions.
Moreover, not all possible structures one might like to construct are included in \cite{Hart-Picciotto-ZomeGeometry}, even when restricted to convex uniform polytopes.
A list available from David Richter's website \cite{Richter-web-page-H4-polychora},  shows real-world constructions of all the $H_4$ polytopes.
At the time of writing, only the bitruncated 120/600-cell has not been constructed according to this page.

This issue highlights the need for a good computational framework. The only existing tool known to the authors for working with Zometool on a computer is Scott Vorthmann's \emph{vZome} software \cite{vzome}, which can be used to construct and render models. vZome is very effective at constructing brand new solids from scratch, and at working with smaller projects, such as the regular polyhedra. For larger tasks, such as projections of 4-dimensional polytopes, it is more difficult to create a vZome construction. While vZome does support its own scripting language, Zomic, it does not possess all the analytic capabilities offered by Maple, and we found some operations we wanted to perform were not supported. 

Another problem we concern ourselves with is the case when structures are {\it not} constructible in Zometool.
When constructing things by hand or in the vZome software, it can be difficult to discern when a structure simply cannot be represented in Zometool, or if we are missing an idea in our construction.

A final issue with existing techniques is the need to be able to zoom in on parts of a model, and look at them in isolation.
The approach used to construct a cube using 20 Zometool pieces is necessarily different from the one used to build the projection of an omnitruncated 120-cell, requiring 21,360 Zometool pieces.
What is needed is a setting in which models can be broken into small workable components, and assembled easily from them, for example in layers.

When deciding constructability, Maple's symbolic nature is desirable, as it allows the output of our program to be taken as a formal proof in either case.
Maple objects can be manipulated and broken apart easily, allowing one to construct a polytope by its individual cells.
In this fashion, ``recipes'' for Zometool constructions can be designed in Maple, streamlining the real-world building process.

\begin{figure}[h]\label{fig:120cell H3 plane}
\centering
\includegraphics[height=3cm]{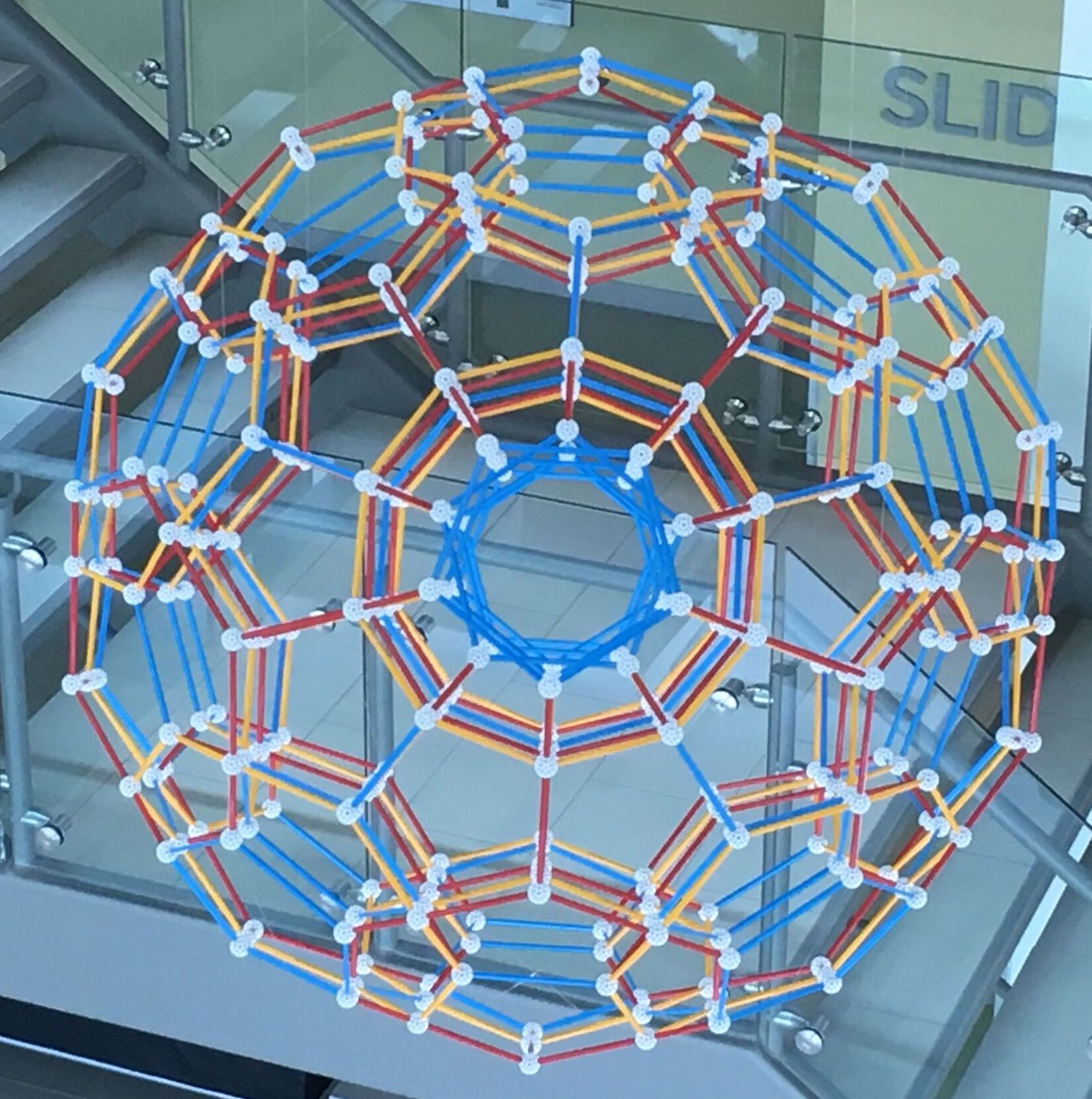}
\hskip1cm
\includegraphics[height=3cm]{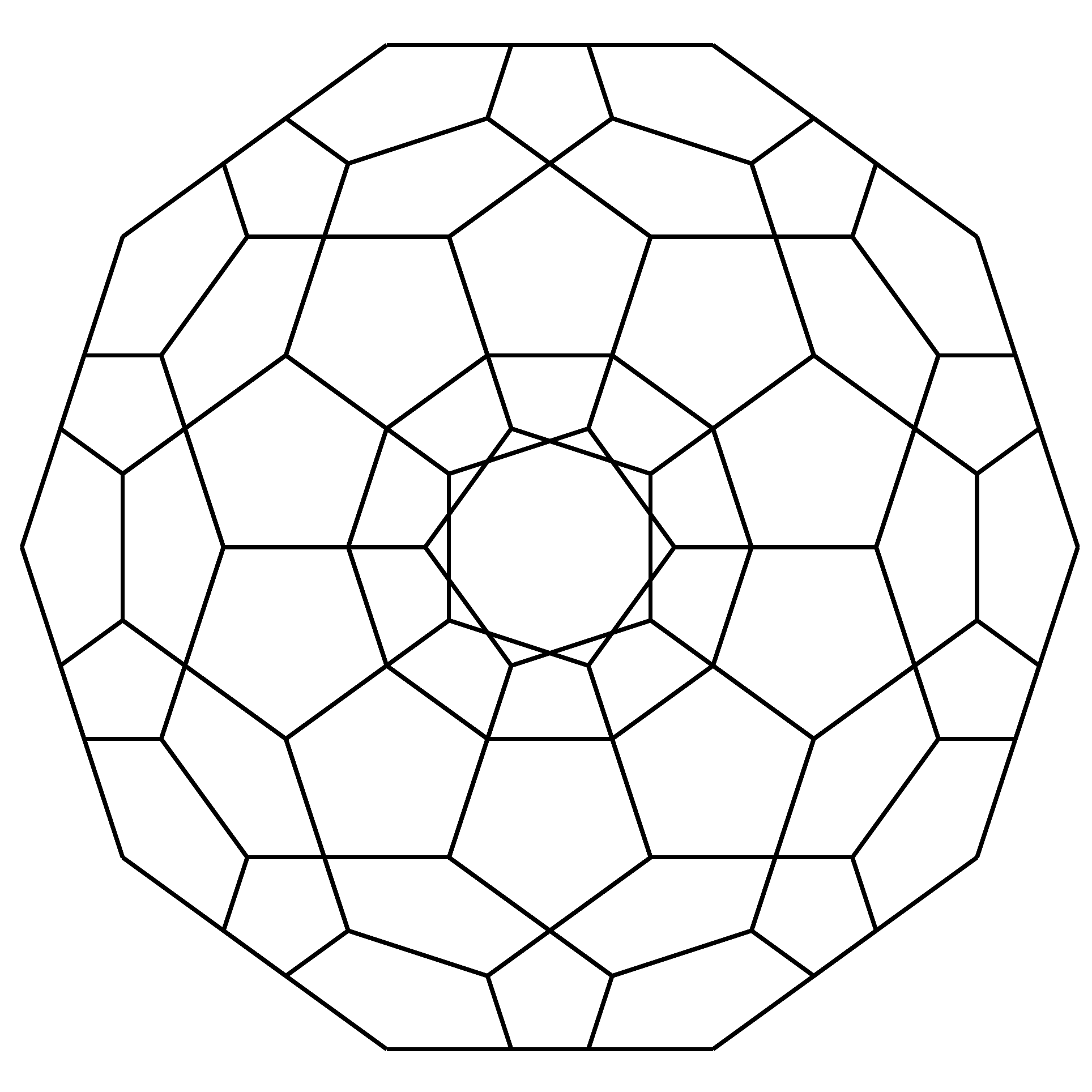}
\caption{The 120-cell projected on the $H_3$ Coxeter plane.}
\end{figure}

One additional note regarding the Wythoffian zoo of examples provided by our package is warranted before we end this introduction. The routines provided allows one to construct all the Wythoffian polytopes, excluding snubs. 
Thus 11 of the 13 Archimedean solids and 45 of the 47 non-prismatic convex uniform 4-dimensional polytopes described in \cite{Conway-Guy-Four-Dimensional-Archimedean} can be generated in this package and tested for Zometool constructability; missing are the snub cube and snub dodecahedron in the Archimedean case, and the snub 24-cell and grand antiprism in the 4-dimensional case.
Finally, for this class of polytopes, we provide functions to perform projections onto Coxeter planes (see Fig.~\ref{fig:120cell H3 plane} for example.)

In Section \ref{sec:withoff_construction}, the Wythoff construction is quickly described. In Section \ref{sec:zome_models}, the possibilities of our package related to Zometool constructions and constructibility are explored. In Section \ref{sec:projections}, the various projections provided by our package are explained and illustrated.
The package is available at \url{https://git.uwaterloo.ca/Zome-Maple/Zome-Maple}.

% section introduction (end)

\section{Wythoff construction} % (fold)
\label{sec:withoff_construction}
To properly discuss this material, a few definitions must be given upfront. A \emph{convex polytope} is a bounded region of $\R^n$ bounded by hyperplanes. The simplest polytopes are thus polygons (in $\R^2$) and polyhedra (in $\R^3$). The intersection of the polytope and its bounding hyperplanes are called \emph{facets}. Hence facets of polygons are edges and facets of polyhedra are faces. A polygon is called \emph{uniform} if it is regular, and a polytope in $n\geq 3$ dimensions is called \emph{uniform} if all of its facets are uniform and the group of symmetry acts transitively on the vertices.

Wythoff had the brilliant idea in his 1918 paper \cite{Wythoff-paper} to construct uniform polytopes by first considering a finite group of reflection acting on $\R^n$ and then considering the polytopes obtained as the convex hull of the orbit of a point under the ion of that group. By construction, these polytopes are uniform. This construction was polished by Coxeter's 1935 paper \cite{Coxeter-WythoffConstruction} following his observation that every finite group of reflections is (what we now call) a Coxeter group: 
\begin{theorem}[Coxeter \cite{Coxeter-DiscreteGroupsGeneratedByReflections}, Theorem 8]
Every finite group of reflections has a presentation of the form $\left\langle r_1,\ldots,r_n\mid (r_ir_j)^{m_{ij}}=1\right\rangle$
for a symmetric matrix $m$ with $m_{ii}=1$. 
\end{theorem}

Coxeter groups are conveniently expressed using \emph{Coxeter diagrams}: labelled graphs whose vertices are the generating reflections $\{r_1,\ldots,r_n\}$ and where the edge $\{r_i,r_j\}$ is labelled $m_{ij}$ and exists only when $m_{ij}\geq 3$. 
\begin{theorem}[Coxeter \cite{Coxeter-FiniteReflectionGroups}]
	Every Coxeter diagram is a finite disjoint union of the Coxeter diagrams for the Coxeter groups $A_n,B_n,D_n,E_6,E_7,E_8,F_4,H_3,H_4$, and $I_2(p)$.
\end{theorem}
In this short paper, we won't describe all these graphs. They are classic and easily found. The upshot of this construction is that it is very easy to describe a polytope using a diagram. The recipe is easy, and one can compute easily the corresponding sub-objects recursively by following the simple algorithm explained in \cite{Champagne1995}. We defer the full explanation to this paper or to Coxeter's book \cite{Coxeter-polytopes} and simply illustrate it by few example.

In dimension three, imagine 3 mirrors passing through the origin and with dihedral angles $\frac\pi2,\frac\pi3,\frac\pi5$. Reflections in this mirror generate a finite group called $H_3$. This configuration is illustrated by the Coxeter diagram \Hthree{}{}{}.
The point chosen is encoded by crossing boxes corresponding to mirror fixing the point. For instance, 
\raisebox{-3pt}{\Hthree{\times}{\times}{}}
 is the polyhedra obtained by taking the convex hull of the orbit under the group $H_3$ of a point at distance 1 from the origin and on the intersection line of the first and second mirror: the icosahedron.  This construction is implemented in Jeff Weeks's \emph{KaleidoTile} software \cite{KaleidoTile} with a visual interface allowing the choice of the point being reflected.
			
KaleidoTile does not allow one to compute the corresponding Wythoffian polytopes in higher dimension. Our package does, and allows one, for example, to compute vertices and edges of the 120-cell \raisebox{-3pt}{\Hfour{}{\times}{\times}{\times}}, the 600-cell \raisebox{-3pt}{\Hfour{\times}{\times}{\times}{}}, and the more complicated omnitruncated 120-cell \Hfour{}{}{}{}.

			% section withoff_construction (end)

\section{Zometool Models} % (fold)
\label{sec:zome_models}
The package uses Maple's module system to work at three distinct layers of abstraction.
The most basic data that is used is the vertices, provided as a list of $n$-tuples, and the edges, a list of unordered pairs of vertices. 
One layer higher is the cell data: Maple's {\tt ComputationalGeometry} package is used to convert the given skeleton into a list of 4-dimensional cells, which may be projected into 3-dimensional space via a function in the package.

\begin{figure}[h]
\centering
\includegraphics[height=4cm]{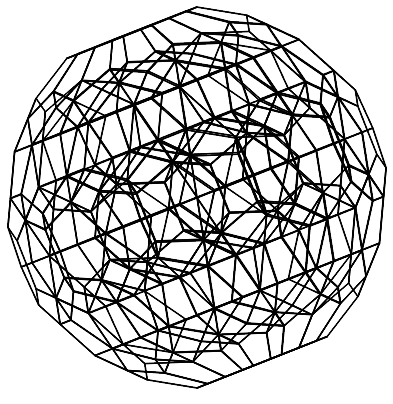}
\hskip1cm
\includegraphics[height=4cm]{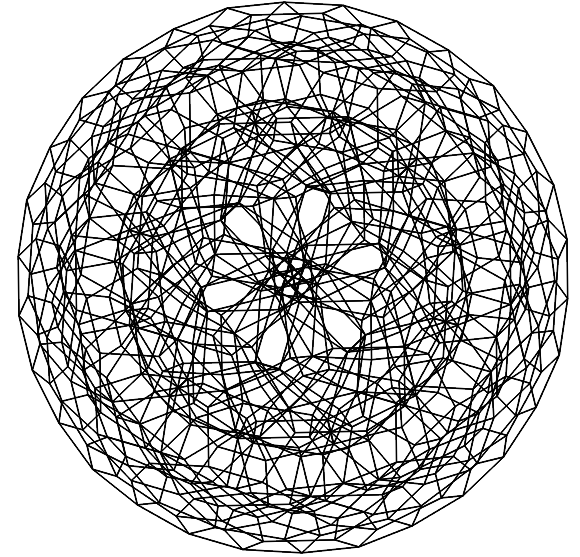}
\caption{Vertex-first and edge-first projections of the 120-cell}
\label{fig:vertexfirst120cells}
\end{figure}

One feature is used to determine whether or not a model is Zometool constructible. For instance, taking the 120-cell and projecting vertex-first to $\R^3$, one finds a set of (normalized) edge lengths not compatible with the Zometool system. This set provides a certificate that this particular projection is not Zometool constructible. 
Regardless of constructibility, our package provides the ability to manipulate the object and display it if desired; see for instance Fig.~\ref{fig:vertexfirst120cells}. 

\begin{figure}
\begin{center}
\includegraphics[height=5.5cm]{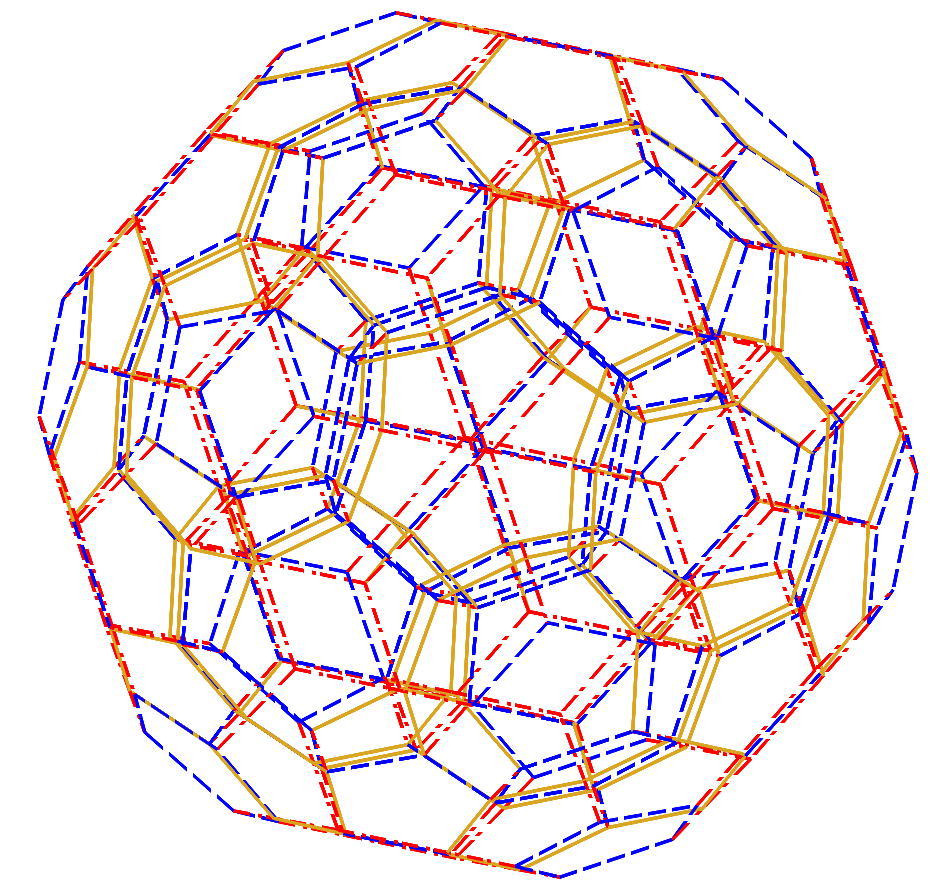}
\hskip5mm
\includegraphics[height=5.5cm]{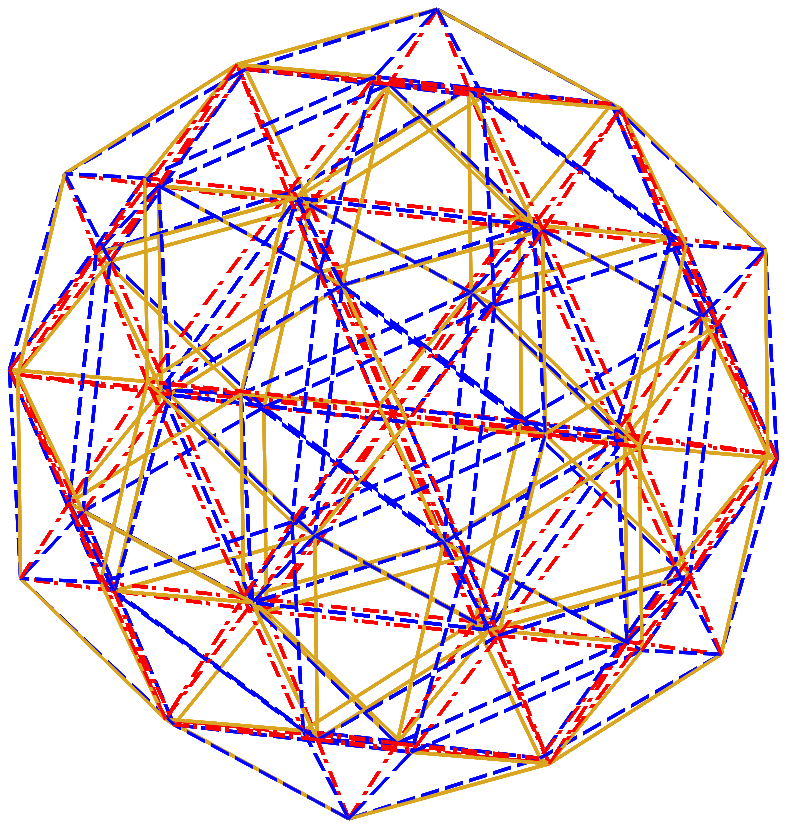}	
\end{center}
\caption{The 120-cell (left) and 600-cell (right) projected cell-first and modeled in Maple. The view is from the $B_3$ and $H_3$ Coxeter planes respectively, and offset slightly to show 3D structure. A dashed line indicates a blue strut, a solid line a yellow strut, and alternating dashes and dots a red strut.}
\label{fig:c120c600}
\end{figure}

\begin{figure}
\centering
\includegraphics[height=3cm]{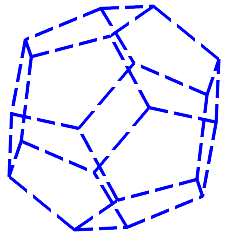}
\hskip1cm
\includegraphics[height=3cm]{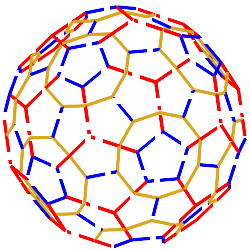}
\caption{Various components of the (cell-first projected) 120-cell. The core  (left), and the upper half of the boundary (right).}
\label{fig:core}
\end{figure}

Assuming the object is Zometool constructible, a list of projected cells can be assembled into a {\it Zometool model}.
Fig.~\ref{fig:c120c600} shows the 120-cell and 600-cell drawn as Zometool models, each projected into three dimensions cell-first.
On its own, these image are too complicated to be of any use, although one point of interest is that they certify the fact that the 120-cell and 600-cell are constructible by Zometool.
We can break apart the image by levels, for example to view the ``core'' of the model, or only the outermost cells, as in Fig.~\ref{fig:core}.

% 120-cell broken by core, equator cells
\begin{figure}
\centering
\includegraphics[height=3cm]{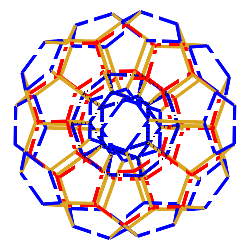}
\hskip1cm %CHANGED: instead of hfill, used hskip with a small distance
\includegraphics[height=3cm]{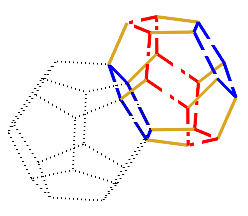}
\caption{The second layer of cells in a (cell-first projected) 120-cell. The type of cell added around the core is shown in the center. The right is the core with only one cell added.}
\label{fig:secondlayer}
\end{figure}

After constructing the core, we can begin using some of the packaged utilities to determine what ought to be built next.
For small models such as the 120-cell, building radially outwards is a standard strategy.
Breaking by levels, we will show the model with its next layer of cells added.
For convenience, the central part is drawn dotted, so the coloured edges are exactly the ones that must be added to the model.
The cell can also be broken off entirely, so that it may be constructed on its own, or shown as the only cell in its layer adjoined to the previous layers.
This makes the picture much less cluttered.
An example is shown in Fig.~\ref{fig:secondlayer}.

%\begin{verbatim}
%> PartialDraw(C120Model, PointNorm, C120Model:-Cells[2])
%> Zome:-ZomeDisplay(ZomeModel(C120Model:-Cells[2], Zome))
%\end{verbatim}

% 120-cell with some more pieces, and the cell that we have to add beside

%To draw a single cell as a Zome model, we extract it with {\tt C120Model:-Cells[2]}, and then pass it to {\tt ZomeModel} to construct a model from it.
%The second argument to {\tt ZomeModel} is the Zome system to be used.
%By varying this argument, Zome systems with different kinds of symmetry can be constructed; for example, a modified Zome system having tetrahedral or octahedral symmetry would be simple to add.
%The default (icosahedral) system is defined in {\tt StandardZome.mla}
%The function {\tt PointNorm} computes the standard Euclidean norm.
%When cells are sorted by the Euclidean norm of their center of mass, the model is built radially outwards.
%The last argument to {\tt PartialDraw} is the ``last'' cell we want to be drawn.
%In this case, it displays all cells with center of mass having norm at most that of {\tt Cells[2]}.

\begin{figure}[h!]
\centering
\includegraphics[height=3cm]{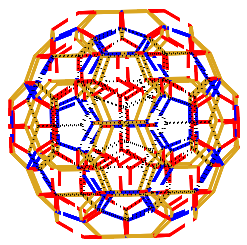}
\hskip1cm
\includegraphics[height=3cm]{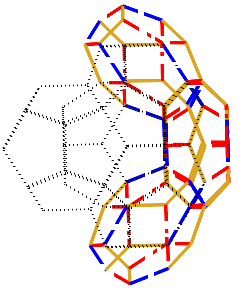}
\hskip1cm
\includegraphics[height=3cm]{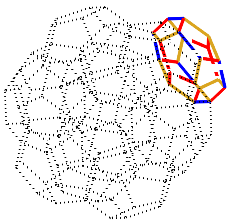}
\caption{The third layer of cells in a 120-cell. In the center is a piece cut from the left model using a filter. The right shows the previous step with only one cell added.}
\label{fig:thirdlayer}
\end{figure}

We can continue this process for two more steps to get the full model.
One useful feature is the ability to pass a filter function; for example, to cut away the cells in all but the positive orthant.
When loaded in Maple, rotating the models is possible, making it somewhat easier to work with than static images.
Using these two tools judiciously together allows one to effectively work with otherwise complicated models.
The next step of the process looks like, with and without a filter function applied, is shown in Fig.~\ref{fig:thirdlayer}.

%\begin{verbatim}
%> PartialDraw(C120Model, PointNorm, C120Model:-Cells[-5])
%> PartialDraw(C120Model, PointNorm, C120Model:-Cells[-5], x ->
%                             evalb(x[1] >= 0 and x[2] >= 0 and x[3] >= 0))
%\end{verbatim}

\begin{figure}[h]
\centering
\includegraphics[scale=0.2]{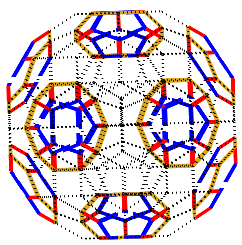}
\hskip1cm
\includegraphics[scale=0.2]{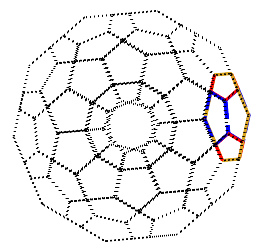}
\caption{Closing off the remaining cells with blue and red pieces completes the model of the 120-cell}
\label{fig:lastlayer}
\end{figure}

Since the 120-cell is uniform, if we understand how to build one part of the layer, we can repeat the construction elsewhere to finish it.
Otherwise, we would have to be more careful with our filters, and handle each part of the layer individually.
Finally, we can close up the last cells to get the full model, as seen in Fig.~\ref{fig:lastlayer}.

%\begin{verbatim}
%> PartialDraw(C120Model, PointNorm, C120Model:-Cells[-1])
%\end{verbatim}

\begin{figure}
\centering
\includegraphics[height=4cm]{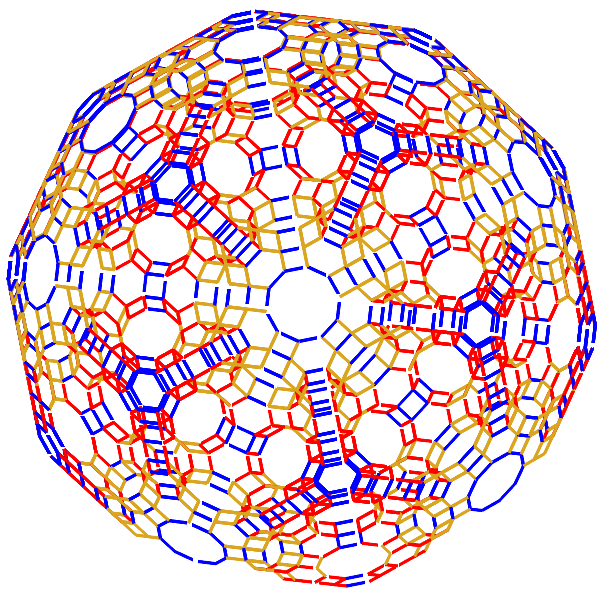}\hskip1cm
\includegraphics[height=4cm]{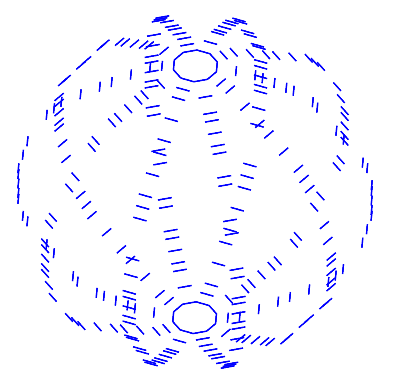}
\caption{Half of the boundary cells of the omnitruncated 120-cell (left), and the four ``blue paths'' in the omnitruncated 120-cell that occur on circles of constant longitude (right).}
\label{fig:omni}
\end{figure}

%More generally, there is a {\tt BreakCellsByLevels} function, which splits cells by the radius from the center they occur at.
%This eliminates the need to guess which cell we work up to in {\tt PartialDraw}.
When considering large models such as the omnitruncated 120-cell, this sort of manipulation is quite helpful.
Since it is impossible to build small-scale physical copies of the model that can be disassembled and investigated (even the smallest incarnation possible in Zometool measures roughly 1.9m in diameter and requires 21,360 pieces), the ability provided by this package to pick apart local features of the overall model is valuable for understanding how it should be constructed.
For example, in order to understand how to suspend the model of the omnitruncated 120-cell from strings, we need to understand what the bottom half of the exterior looks like, to decide where strings should be placed.
Some special paths formed by blue edges make good candidates for these string paths, and one could want to be able to isolate this feature.
The results of both of these computations are shown in Fig.~\ref{fig:omni}.

These are not all the operations supported by the library, and generally it is easy to extend it to perform any other specific manipulations you might need.
What we are trying to show is that by casting the question of Zometool modelling in the established framework of Maple, we get access to a powerful set of tools that can help in many aspects of a large-scale Zometool project.

One final use of this package that we shall point out is the ability to generate parts lists for a model.
This is a rather long computation to run by hand, but simple to compute in Maple.
Here is the list generated for the omnitruncated 120-cell, projected through a great rhombicosidodecahedral cell.

\begin{verbatim}
Balls = 7200
R2 = 2880
R1 = 2880
B2 = 3600
Y2 = 4800
\end{verbatim}

Computations of this sort allow us to verify entries in Richter's list \cite{Richter-zome-600-cell}, for example. 
% section zome_models (end)

\section{Projections} % (fold)
\label{sec:projections}

\begin{figure}[h]
\centering
\includegraphics[height=5cm]{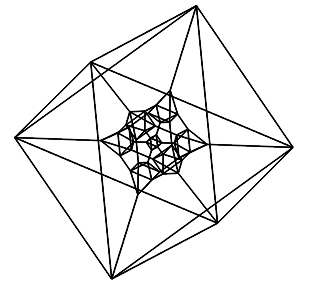}\hskip5mm\includegraphics[height=5cm]{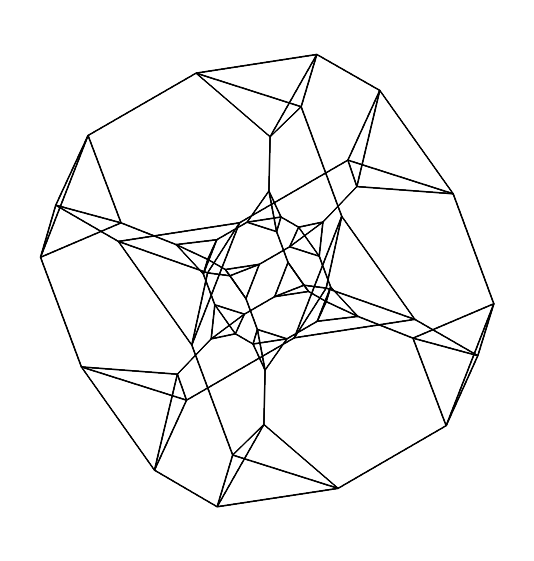}
\caption{A north pole stereographic projection of the truncated 16-cell  and truncated hypercube.}
\label{fig:stereo}
\end{figure}

In addition to constructing Zometool models, once cell data is constructed, it can be projected into three or two dimensional space and drawn, regardless of Zometool construtability.
The projections included in this package are orthogonal projections onto arbitrary bases, stereographic projections, and projections onto Coxeter planes.
When creating Zometool constructions of 4-dimensional polytopes, the most useful of these is the orthogonal projection---stereographic projections distort distances too much, and Coxeter plane projections are two-dimensional.
We include the other projections, as in many cases they are not available elsewhere, in some cases are very nice looking, and in general are useful to get a good grasp of the objects; see for instance in Fig.~\ref{fig:stereo} the stereographic projections of the truncated 16-cell (\FourNodesDiagram{4}{}{}{\times}{\times}{}{}) and truncated hypercube (\FourNodesDiagram{4}{}{}{}{}{\times}{\times}).

\begin{figure}[h]
\centering
\includegraphics[height=5cm,trim = 100 0 100 0,clip=true]{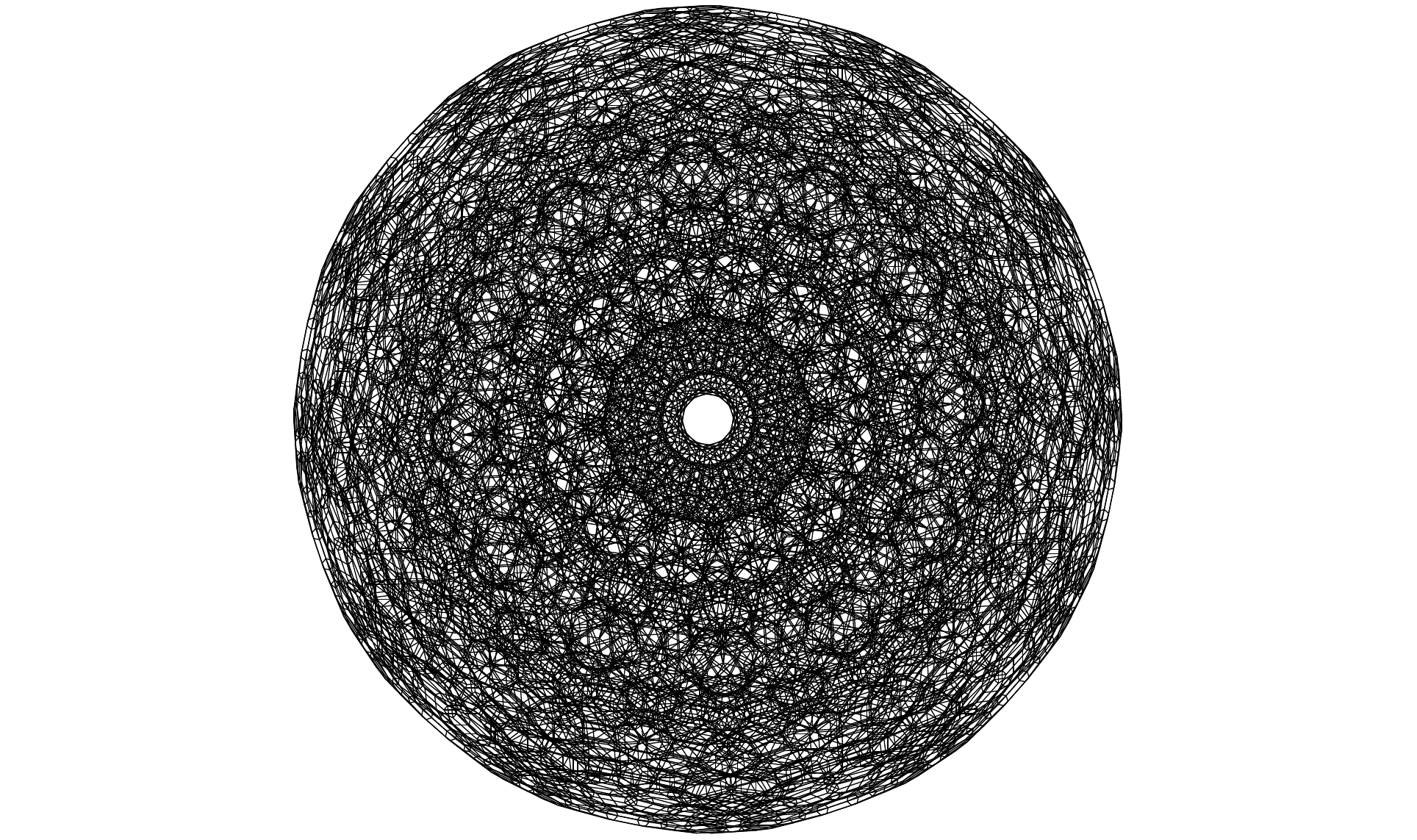} 
\includegraphics[height=5cm,trim = 100 0 100 0,clip=true]{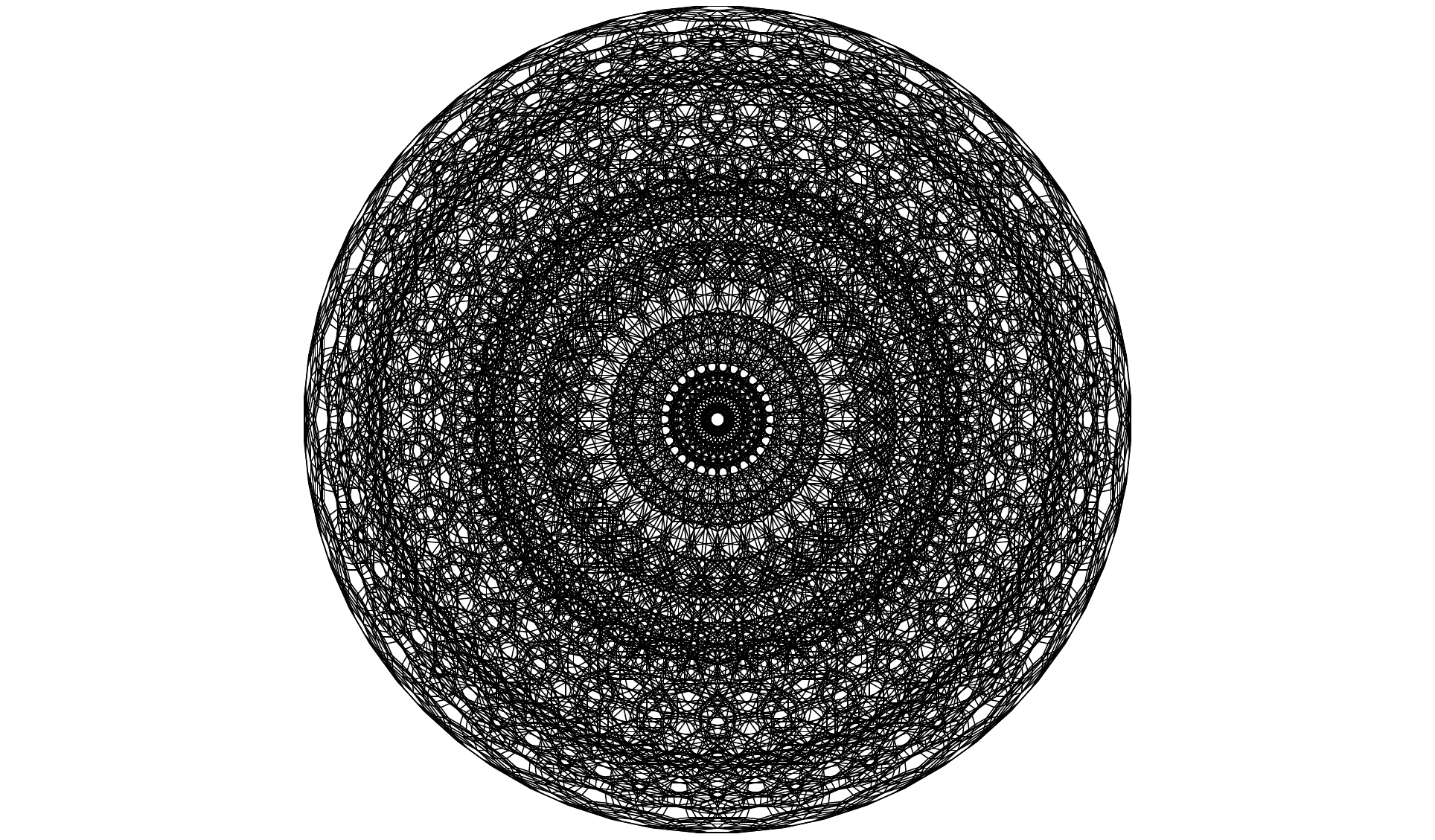}
%TODO: remove the extra white column in those eps
\caption{The $F_4$ and $H_4$ Coxeter plane projection of the omnitruncated 120-cell}
\label{fig:omnih4}
\end{figure}

Of particular interest are the vertex-first, edge-first, face-first, cell-first projections, see for instance Fig.~\ref{fig:vertexfirst120cells}.
Shown in Fig.~\ref{fig:omnih4} are the $F_4$ and $H_4$ projections of the omnitruncated 120-cell computed by our package.

These projections do not occur as projections of the Zometool model of the omnitruncated 120-cell.
So, in addition to allowing us to study Zometool models of Wythoffian polytopes, the Coxeter plane projections in this package can be used to enjoy some of the higher dimensional structure that is lost when projecting into three dimensions for the purposes of Zometool construction.

% \begin{figure}[h!]
% \centering
% \includegraphics[height=5cm]{COmni-F4}
% \caption{The $F_4$ Coxeter plane projection of the omnitruncated 120-cell}
% \label{fig:omnif4}
% \end{figure}

% section projections (end)

 \bibliographystyle{splncs04}
 \bibliography{ZomeGeometry,../polytopeLiterature/polytopes}

\begin{thebibliography}{10}
\providecommand{\url}[1]{\texttt{#1}}
\providecommand{\urlprefix}{URL }
\providecommand{\doi}[1]{https://doi.org/#1}

\bibitem{Zometool-about}
A (not so brief) history of {Zometool},
  \url{https://www.zometool.com/about-us/}

\bibitem{Champagne1995}
Champagne, B., Kjiri, M., Patera, J., Sharp, R.T.: Description of
  reflection-generated polytopes using decorated {Coxeter} diagrams. Canadian
  J. Physics  \textbf{73},  566--584 (1995). \doi{10.1139/p95-084}

\bibitem{Chuang-Jin-Superfullerenes-Zome}
Chuang, C., Jin, B.Y.: Construction of {Sierpi\'nski }superfullerenes with the
  aid of {Zome} geometry: Application to beaded molecules. In: Hart, G.W.,
  Sarhangi, R. (eds.) Proceedings of Bridges 2013: Mathematics, Music, Art,
  Architecture, Culture. pp. 495--498. Tessellations Publishing, Phoenix,
  Arizona (2013),
  \url{http://archive.bridgesmathart.org/2013/bridges2013-495.html}

\bibitem{Coxeter-DiscreteGroupsGeneratedByReflections}
Coxeter, H.S.M.: Discrete groups generated by reflections. Annals of
  Mathematics  \textbf{35}(3),  588--621 (1934),
  \url{http://www.jstor.org/stable/1968753}

\bibitem{Coxeter-FiniteReflectionGroups}
Coxeter, H.S.M.: The complete enumeration of finite groups of the form $r_i^2 =
  (r_i r_j)^{k_{ij}} = 1$. J. London Math. Soc., (1)  \textbf{10},  21--25
  (1935). \doi{10.1112/jlms/s1-10.37.21}

\bibitem{Coxeter-WythoffConstruction}
Coxeter, H.S.M.: Wythoff's {C}onstruction for {U}niform {P}olytopes. Proc.
  London Math. Soc. (2)  \textbf{38},  327--339 (1935).
  \doi{10.1112/plms/s2-38.1.327}

\bibitem{Coxeter-polytopes}
Coxeter, H.S.M.: Regular polytopes. Dover Publications Inc., New York, third
  edn. (1973)

\bibitem{Conway-Guy-Four-Dimensional-Archimedean}
Conway, J.H., Guy, M.J.T.: Four-dimensional {Archimedean} polytopes. In:
  Proceedings of the Colloquium on Convexity, Copenhagen, 1965. K{\o}benhavns
  Universitets Matematiske Institut, Copenhagen, Denmark (1965)

\bibitem{Hall-textbook}
Hall, B.: Lie groups, {L}ie algebras, and representations, Graduate Texts in
  Mathematics, vol.~222. Springer, Cham, second edn. (2015).
  \doi{10.1007/978-3-319-13467-3}, an elementary introduction

\bibitem{Hall-GeometryRootSystems}
Hall, B.C.: The geometry of root systems: an exploration in the {Zometool}
  system, \url{https://www3.nd.edu/~bhall/book/lie.htm}

\bibitem{Hart-zome-polytopes-BarnRaising}
Hart, G.W.: Barn raisings of four-dimensional polytope projections. Proceedings
  of International Society of Art, Math, and Architecture 2007  (2007),
  \url{http://www.georgehart.com/zome-polytopes-ISAMA07/hart-zome-polytopes.pdf}

\bibitem{Hart-Picciotto-ZomeGeometry}
Hart, G.W., Picciotto, H.: Zome geometry: Hands-on learning with {Zome} models.
  Key Curriculum Press (2001)

\bibitem{Hildebrandt-Zome-workshop}
Hildebrandt, P.: Zome workshop. In: Sarhangi, R., Barrallo, J. (eds.) Bridges
  Donostia: Mathematics, Music, Art, Architecture, Culture. pp. 459--464.
  Tarquin Publications, London (2007),
  \url{http://archive.bridgesmathart.org/2007/bridges2007-459.html}

\bibitem{Richter-web-page-H4-polychora}
Richter, D.A.: H(4)-polychora with {Zome},
  \url{http://homepages.wmich.edu/~drichter/h4polychorazome.htm}

\bibitem{Richter-zome-600-cell}
Richter, D.A.: Two results concerning the {Zome} model of the 600-cell. In:
  Sarhangi, R., Moody, R.V. (eds.) Renaissance Banff: Mathematics, Music, Art,
  Culture. pp. 419--426. Bridges Conference, Southwestern College, Winfield,
  Kansas (2005),
  \url{http://archive.bridgesmathart.org/2005/bridges2005-419.html}

\bibitem{Richter-Vorthman-Green-Quaternions-OctahedealZome}
Richter, D.A., Vorthmann, S.: Green quaternions, tenacious symmetry, and
  octahedreal {Zome}. In: Sarhangi, R., Sharp, J. (eds.) Bridges London:
  Mathematics, Music, Art, Architecture, Culture. pp. 429--436. Tarquin
  Publications, London (2006),
  \url{http://archive.bridgesmathart.org/2006/bridges2006-429.html}

\bibitem{Voros-Zometool-B-DNA}
V\"or\"os, L.: A {Zometool} model of the b-dna. In: Torrence, E., Torrence, B.,
  S\'equin, C., McKenna, D., Fenyvesi, K., Sarhangi, R. (eds.) Proceedings of
  Bridges 2016: Mathematics, Music, Art, Architecture, Education, Culture. pp.
  435--438. Tessellations Publishing, Phoenix, Arizona (2016),
  \url{http://archive.bridgesmathart.org/2016/bridges2016-435.html}

\bibitem{vzome}
Vorthmann, S.: {vZome}, software available at \url{http://vzome.com/home/}

\bibitem{KaleidoTile}
Weeks, J.: {KaleidoTile}, software available at
  \url{http://geometrygames.org/KaleidoTile/index.html.en}

\bibitem{Wythoff-paper}
{Wijthoff}, W.A.: {A relation between the polytopes of the {C600}-family}.
  Koninklijke Nederlandse Akademie van Wetenschappen Proceedings Series B
  Physical Sciences  \textbf{20},  966--970 (1918)

\end{thebibliography}

\end{document}